\documentclass[twoside]{article}
\usepackage{fleqn,espcrc2}

\newcommand{\bb}{\begin{eqnarray}}
\newcommand{\ee}{\end{eqnarray}}

\newcommand{\AmS}{{\protect\the\textfont2
  A\kern-.1667em\lower.5ex\hbox{M}\kern-.125emS}}

\title{Theories with global gauge anomalies on the lattice}

\author{P. Mitra\address{Saha Institute of Nuclear Physics,
Block AF, Bidhannagar, Calcutta 700 064, INDIA}
\thanks{e-mail mitra@tnp.saha.ernet.in}}
       
\begin{document}

\begin{abstract}
	A global anomaly in a chiral gauge theory manifests itself in
different ways in the continuum and on the lattice. In the continuum case,
functional integration of the fermion determinant over the whole space of
gauge fields yields zero. In the case of the lattice, it is not even
possible to define a fermion measure over the whole space of gauge
configurations. However, this is not necessary, and as in the continuum, a
reduced functional integral is sufficient for the existence of the theory.

\vspace{1pc}

\centerline{\bf  Presented at Lattice 2000, Bangalore \quad hep-lat/0010003}
\end{abstract}

\maketitle

\section{Introduction}

Anomalies  are of  two  different  types.
Local or divergence anomalies have been known since 1969 \cite{1}: 
%the effective action has a reduced symmetry and
classically conserved symmetry currents 
cease to be conserved after quantization if there are anomalies of this kind.
For example, the theory
\bb
{\cal L}=\bar\psi(i\partial\!\!\!{/} -eA\!\!\!{/})\psi
-\frac{1}{4}F_{\mu\nu}F^{\mu\nu}\ee
does not obey the expected conservation law for its axial current:
\bb
\partial_\mu(\bar\psi\gamma^\mu\gamma_5\psi)=
{e^2\over 16\pi^2}F_{\mu\nu}F^{\mu\nu}\neq 0.\ee
If an anomalous current is associated with a {\em gauged symmetry}, 
it leads to an
apparent problem in quantization because
the equations of motion of
the gauge  fields  require the   current   to   be   conserved.
A treatment of such a theory just like a usual gauge theory shows an
inconsistency.
This problem can be sorted out by paying proper attention to
phase space constraints, as suggested by \cite{Faddeev}.
The anomaly itself can be made  to  vanish  in a sense
by going to  the constrained submanifold of classical
phase space. However,
theories  with  anomalous gauge  currents are to be distinguished from
theories  with  nonanomalous gauge  currents. 
If a theory is {\em nonanomalous}, it possesses gauge freedom, 
and is describable in any one of an infinite
variety  of  gauges,  
whereas an {\em anomalous gauge
theory} may have its gauge fixed by the anomaly itself. 

The second  kind  of  anomaly is the  
{\em topological or global} anomaly  discovered in 1982 \cite{Witten}.
The gauge current is conserved here, but  
the topology of the gauge group in the continuum is such that the
fermion determinant changes 
under {\em large} gauge transformations
{\it i.e.,} those not continuously connected to the identity transformation.
This may result in a {\em vanishing of the partition function}.
An example is provided by the SU(2) gauge theory with a doublet of 
Weyl fermions.
Here gauge transformations fall into two disconnected classes, those connected
to the identity, and the others.
The fermion determinant picks up a
change of sign if the gauge field in which it is evaluated is
transported along a (non-gauge) path in the gauge configuration space
connecting two large-gauge-transformation-related configurations.
Such theories nevertheless exist in the continuum, as shown in 
\cite{Mitra} by appealing to canonical quantization.

After the recent developments around
the Ginsparg-Wilson relation for lattice fermions,
a general formulation
of global anomalies on the lattice has been given \cite{Baer}.
Our aim here is to show that the theories exist on the lattice as 
well, though an explicit construction cannot be given at the moment.

The continuum argument of \cite{Mitra} is recapitulated in the next
section. Then we pass to the lattice.

\section{Functional integral in the continuum}

The full partition function of 
the gauge theory with fermions may be written as 
\bb Z=\int {\cal D}A Z[A],\ee
\bb Z[A]\equiv e^{-S_{eff}}=\int {\cal D}\bar\psi {\cal D}\psi  
e^{-S(\psi,\bar\psi,A)}\ee
An anomaly-free theory has $Z[A]$ gauge invariant.
If there is a gauge anomaly,  $Z[A]$ varies under gauge transformations of $A$: 
\bb Z[A^g]=e^{i\alpha (A, g^{-1})} Z[A],\ee 
where $\alpha$ may be regarded as an integral representation  of the anomaly.
It obeys some consistency conditions (mod $2\pi$):
\bb \alpha(A, g_2^{-1}g_1^{-1})&=&\alpha(A, g_1^{-1})+\alpha(A^{g_1},
g_2^{-1})\nonumber\\
\alpha(A, g^{-1})&=&-\alpha(A^g,g).\ee
The case becomes one of a global anomaly if  $\alpha$ is independent of $A$,  
and vanishes for $g$
connected  to  the  identity but not for some $g$ which cannot be continuously
connected to the identity.
$A$-independence implies an abelian representation satisfying
\bb \alpha(g_2g_1)=\alpha(g_1)+\alpha(g_2).\ee
In the SU(2) case, the two components of the gauge group manifest themselves
in two possible values of the phase: $e^{i\alpha}=\pm 1$.

In an anomaly-free theory, the partition function factorizes
into the volume of the gauge group and the gauge-fixed partition function:
\bb Z&=&\int {\cal D}AZ[A]\nonumber\\
&=&\int {\cal D}AZ[A]\int {\cal D}g\delta(f(A^g))\Delta_f(A)\nonumber\\
&=&\int {\cal D}g\int {\cal D}AZ[A^{g^{-1}}]\delta(f(A))\Delta_f(A)\nonumber\\
&=&\int {\cal D}g\int {\cal D}AZ[A]\delta(f(A))\Delta_f(A)\nonumber\\
&=&(\int {\cal D}g)Z_f\ee
This is the standard Faddeev-Popov argument.
Here,  
$\delta(f)$ represents a gauge-fixing operation and
$\Delta_f$ is the corresponding Faddeev-Popov determinant. 
%(Note: in deriving the fourth equality, use is made of the 
%invariance of $Z[A]$ under gauge transformations.)

This  decoupling of gauge degrees of freedom does not occur
if a local anomaly is present.  
%\bb Z=\int {\cal D}g\int {\cal D}A
%e^{i\alpha(A,g)}Z[A]\delta(f(A))\Delta_f(A),\ee
%but $g$ and $A$ are coupled by the anomaly
%term $\alpha(A,g)$. 
%
For a global anomaly however, the partition function
factorizes again:
\bb Z=\int {\cal D}g
e^{-i\alpha(g)}\int {\cal D}AZ[A]\delta(f(A))\Delta_f(A)\ee
As the phase factors form  a representation
of the gauge group, 
\bb \int {\cal D}ge^{-i\alpha(g)}&=&
\int {\cal D}(gh)e^{-i\alpha(gh)}\nonumber\\ &=& 
e^{-i\alpha(h)} \int {\cal D}ge^{-i\alpha(g)}\ee
where $h$ stands for a fixed gauge transformation. If $h$
is {\em not} connected to the identity, 
$e^{-i\alpha(h)}\neq 1$, and consequently  
$\int {\cal D}ge^{-i\alpha(g)}=0$, which in turn means that
$Z=0$.
Does this mean that the theory cannot be defined? Let us look at 
expectation values of gauge invariant operators.
\bb &&{\int{\cal D}AZ[A]{\cal O}\over\int{\cal D}AZ[A]}=\nonumber\\
&&{\int {\cal D}ge^{-i\alpha(g)}
\int{\cal D}AZ[A]\delta(f(A))\Delta_f(A){\cal O}
\over\int {\cal D}ge^{-i\alpha(g)}
\int {\cal D}AZ[A]\delta(f(A))\Delta_f(A)}\nonumber\\
&& \ee
The expression on the right is of the form ${0\over 0}$ because the 
factor $ \int {\cal D}ge^{-i\alpha(g)}$ vanishes, as we have seen above. 
Can one not reinterpret the ratio  by removing this common vanishing
factor? 
\bb <{\cal O}>\stackrel{?}{=}
{\int{\cal D}AZ[A]\delta(f(A))\Delta_f(A){\cal O}
\over \int {\cal D}AZ[A]\delta(f(A))\Delta_f(A)}.\ee

The right hand side is precisely what one gets in the {\em canonical} approach
to quantization where gauge degrees of freedom are removed by fixing the gauge
at the classical level and only physical degrees of freedom enter the
functional integral.  The Faddeev-Popov determinant arises in the canonical
approach as the determinant of the matrix of Poisson brackets of what may be
called the "second class constraints", {\it i.e.}, the Gauss law operator and
the gauge fixing condition $f$, which is of course introduced by hand and not
really a constraint of the theory.  There are both ordinary fields and
conjugate momenta, but the latter are easily integrated over.  The point is
that the full functional integral is not needed in the canonical approach and
there is no harm if it vanishes!

A  trace is left behind by the global anomaly. 
One may imagine a classification of the gauge-fixing  functions  $f$
where $f, f'$ are said to belong to the same  class
if there exists a  gauge  transformation  connected to 
the identity to go from a configuration with $f=0$
to one with $f'=0$. Then
$Z_f=Z_{f'}.$
More generally, when such a transformation is not connected
to the identity,
%\bb && Z_f\nonumber\\ 
%&&= \int {\cal D}AZ[A]\delta(f(A))\Delta_f(A)\nonumber\\ 
%&&= \int {\cal D}AZ[A]\delta(f(A))\Delta_f(A)\cdot\nonumber\\
%&&\int {\cal D}g\delta(f'(A^g))\Delta_{f'}(A)\nonumber\\
%&&=\int{\cal D}g\int {\cal D}AZ[A]\delta(f(A))\cdot\nonumber\\
%&&\Delta_f(A) \delta(f'(A^g))\Delta_{f'}(A)\nonumber\\
%&&=\int{\cal D}g\int {\cal D}AZ[A^{g^{-1}}]\delta(f(A^{g^{-1}}))\Delta_f(A)
%\cdot\nonumber\\
%&&\delta(f'(A))\Delta_{f'}(A) \nonumber\\ 
%&&= \int {\cal D}AZ[A]\delta(f'(A))\Delta_{f'}(A)\cdot\nonumber\\
%&&\int {\cal D}ge^{-i\alpha(g)}\delta(f(A^{g^{-1}}))\Delta_f(A)
%\ee 
%If there were no anomaly, $e^{-i\alpha(g)}=1$, and the last integral would be equal to
%unity and the right hand side would reduce to $Z_{f'}$.
%Even with a global anomaly, a decoupling occurs: 
%Only those configurations are
%relevant for which $f'(A)=f(A^{g^{-1}})=0.$ The consequence is that
%one $g$ is picked out for each $A.$ Now
%$g$ varies in a fixed homotopy class, so $\alpha(g)
%=\alpha(g_0),$ a constant
%depending only on the class:
\bb Z_f=e^{-i\alpha(g_0)}Z_{f'},\ee
where $g_0$ is determined by  $f,f'$. 
These factors $e^{-i\alpha(g_0)}$
occurring in partition functions
cancel out in expectation values of gauge invariant
operators, so that Green  functions  of  gauge  invariant
operators are fully gauge independent \cite{Mitra}.

There is an assumption in all this: that there is a
possibility of fixing the gauge.
A general theorem \cite{Singer} asserts that gauges
{\em  cannot} be fixed in a smooth way. 
For the construction of functional integrals, however, it is 
sufficient to have  {\em piecewise smooth gauges}.
It should also be remembered that these questions arise
even for theories {\em without} disconnected gauge groups and are not specific
to the context of global anomalies.

\section{Lattice formulation}

On going to the lattice, one starts to use group-valued
variables associated with links instead of $A$ defined at points of the continuum.
The topology also changes: the gauge group becomes {\em connected} on the lattice:
it becomes possible to go to any gauge transformation from the identity 
in a continuous manner. Thus there are
no large gauge transformations any more. Does it mean that there is
no global anomaly on the lattice?
The issue is complicated because chiral symmetry is not straightforward here.
Chiral symmetry on the lattice has begun to make more sense in the last few years
thanks to the Ginsparg-Wilson relation
imposed on $D$, the euclidean lattice Dirac operator: 
\bb
\gamma_5D+D\gamma_5=aD\gamma_5D,\ee
where $a$ is the lattice spacing.
An analogue of $\gamma_5$ appears from the above relation:
\bb
\gamma_5D=-D\Gamma_5, ~\Gamma_5\equiv\gamma_5(1-aD).\ee
It satisfies 
\bb
(\Gamma_5)^2=1,~ (\Gamma_5)^\dagger=\Gamma_5,\ee
and can be used to define left-handed projection:
\bb
P_-\psi&\equiv&\frac{1}{2}[1-\Gamma_5]\psi=\psi\nonumber\\
\bar\psi P_+&\equiv&\bar\psi\frac{1}{2}[1+\gamma_5]=\bar\psi.\ee
In this way of defining chiral projections,
$P_-$, but not $P_+$, depends on the gauge field configuration.
Nontriviality of chirality on the lattice stems from this $P_-$.

A fermion measure is defined by specifying a basis of lattice Dirac fields
$v_j(x)$ satisfying 
\bb
P_-v_j=v_j,~ (v_j,v_k)=\delta_{jk}.\ee
One has to integrate over Grassmann-valued expansion coefficients in
\bb
\psi(x)=\sum_j a_jv_j(x).\ee
Expansion coefficients also come from
the expansion of $\bar\psi$ in terms of $\bar v_j$ satisfying
$\bar v_jP_+=\bar v_j$,
but these are as usual, {\it i.e.}, do not involve gauge fields.

Questions of {\em locality} and {\em integrability} arise
because of the gauge field dependence in $P_-$.
Absence of a local anomaly appears to be sufficient to ensure locality \cite{Luesch}.
Global anomalies are manifested as a lack of {\em integrability}.

Consider, following \cite{Baer}, a closed path in the SU(2) gauge
configuration space, with the parameter $t$ running from 0 to 1. Define
\bb
f(t)=\det [1-P_++P_+D(t)Q_t{D(0)}^\dagger],\ee
with $D(t)$ the  Dirac operator corresponding to gauge fields at
parameter value $t$, and
$Q_t$ the unitary transport operator for $P_-(t)$ defined by
\bb
\partial_tQ_t=[\partial_tP_-(t),P_-(t)]Q_t, ~ Q_0=1.\ee
Then $f(t)$ is real, positive and satisfies 
\bb
f(1)={\cal T}f(0).\ee
Here 
\bb
{\cal T}=\det[1-P_-(0)+P_-(0)Q_1]=\pm 1\ee
depending on the topology of the considered path in the gauge configuration space.
$f$ changes sign an even or odd number of times along path depending on ${\cal T}$
and while $\det D(t)$ is related to $f^2$,
\bb 
\det D(t)\det {D(0)}^\dagger= f^2(t),\ee
%$\Rightarrow \det D(t)$ undergoes no change of sign, but
%$\det D$ is of the form $\prod_i \lambda_i\lambda_i^*$, as
%the eigenvalues $\lambda_i$ have to appear in conjugate pairs because of
%the property $D^\dagger=\gamma_5D\gamma_5$, and it is the number of sign 
%changes of the eigenvalues $\lambda_i$ that is of interest because 
the chiral fermion determinant $\det D_\chi(t)$ behaves like $f$: 
\bb
&&\det D_\chi(t)\det D_\chi(0)^\dagger=f(t)W(t)^{-1},\nonumber\\
&&(D_\chi)_{ij}\equiv a^4\sum_x \bar v_i(x)Dv_j(x).\ee  
Here $W(t)$ is a  phase factor arising from the gauge field dependence of $v_j$.
It is a lattice artifact and may be  taken to reduce to unity 
near the continuum limit. 

Then $\det D_\chi$ changes sign, {\it i.e.}, 
fails to return to its starting value after transportation along a closed path
if the path has 
\bb
{\cal T}=-1.\ee
Such paths have been shown to exist in the SU(2) theory. A part of such a path 
lies along a gauge  orbit, and a part is non-gauge.\footnote 
{The integrability condition valid for such paths is ${W(1)={\cal T}}$,
but in the near-continuum region considered in the current literature,
$W=1+{\cal O}(a)$ cannot become -1.}

Thus $\det D_\chi$ is   multivalued, implying that
the fermion measure is not well defined, and hence the
functional integral does not make sense.
This is roughly similar to the continuum.
The Dirac operator is gauge-invariant and its determinant
and $f$ can change only on non-gauge
portions of the closed path. So the problem of sign change of $f$
occurs once again in non-gauge paths connecting gauge-related configurations.
However, in the continuum, the sign change
occurs between configurations which can be 
connected only by a non-gauge path.
On the lattice, the sign change occurs 
when configurations are  connected by  a non-gauge path, though a
connection is also possible by   a gauge path, with no accompanying change of sign. 
In other words, the sign or phase is ambiguous.
There is an obvious remedy: to  restrict the functional integral
to one point on each orbit, {\it i.e.},  to fix the gauge.
Problematic paths with ${\cal T}=-1$ can thereby be avoided. The
fermion measure is defined over the reduced configuration space and the theory  exists.

Note that $W$ need not be taken to be $1+{\cal O}(a)$. This may
yield an alternative way of defining the theory on a finite
lattice.

As in the continuum, there is the question whether the gauge can be fixed.
In the lattice literature, there is a lot of discussion on gauge fixing,
Gribov ambiguities and related phenomena. Without going into these details,
one can see that if one considers a single gauge orbit as a set, it will in
principle be possible to pick a configuration from that set, and this fixes
the gauge for a single orbit. The procedure can be repeated for each orbit.
A problem arises only when one tries to combine these choices for different
orbits. A smooth gauge is not expected to exist, but a piecewise smooth choice
should be possible, as in the continuum. There are hopes of making
gauge choices which lead to BRST invariance in the continuum limit
\cite{Testa}.
   
\section*{Acknowledgment}
I would like to thank Maarten Golterman for some comments on gauge 
fixing on the lattice.


\begin{thebibliography}{99}
\bibitem{1} S. Adler, { Phys. Rev.}  177 (1969) 2426;
J. S. Bell and R. Jackiw, { Nuov. Cim.}  60A (1969) 47.
%\bibitem{7} L. D. Faddeev, { Theo. Math. Phys.} {\bf 1} (1969)1.
\bibitem{Faddeev} L. D. Faddeev, { Phys. Lett.}  145B (1984) 81.
\bibitem{Witten} E. Witten, { Phys. Lett.}  117B (1982) 324.
\bibitem{Mitra} P. Mitra, { Lett. Math. Phys.}  31 (1994) 111; see also
P. Mitra, hep-th/9506158.
\bibitem{Baer} O. B\"{a}r and I. Campos, DESY-99/188.
\bibitem{Singer} I. M. Singer, { Comm. Math. Phys.}  60 (1978) 7.
\bibitem{Luesch} M. L\"{u}scher, hep-lat/9904009.
\bibitem{Testa} M. Testa, hep-lat/9803025.
\end{thebibliography}
\end{document}